# Democratizing information visualization.
# A study to map the value of graphic design to easier knowledge transfer of scientific research.

Matteo Zallio[1]

[1] University of Cambridge, Department of Engineering, Trumpington st., Cambridge, United Kingdom
mz461@cam.ac.uk

**Abstract:**
Visual representations are a consistent component of human evolution. They are forms of concept designs using combinations of colors, patterns, and geometrical figures.
As history displays, graphical visualization has been used since early ages as a mean to transfer knowledge between human beings across generations.
Recently, visual representations became recognized by various scientists as tools to ease cognitive comprehension in various scientific fields. There-fore, visual representations are becoming important in science communication and education.
Notwithstanding the general relevance of visual representations, there is not extensive recognition of the value of graphical representation to im-prove communication of scientific results and the level of awareness of visual design principles among non-design-centered communities is low. Considering these aspects, this explorative study investigates the perception of STEM researchers, without any specific visual design background, and the value of visual representations as tools to support the communication of technical and scientific knowledge among academics and a wider non-design-centered community. To measure the user perception and the value of visual representations in daily working practice, an interpretive approach was used, and a pilot study was conducted with a group of voluntary participants recruited among research members from a well-known American institution.
Results emphasize how researchers, who included professionally executed graphical representations in their publications, reports, and grant proposals, perceived their work as more easily readable. Early findings show that visual representations can positively support scientists to share research out-comes in a more compelling, visually clear, and impactful manner, reaching a wider audience across different disciplines.

## 1. Introduction: The role of visual representations

Visual representation lays the foundation of its origin since the early ages of humans living on Earth. From prehistoric symbols in the Lascaux caves, estimated at around 17,000 years ago [1], to engraved symbols in Rome's Trajan's Column (A.D. 113) [2], to Illuminated manuscripts from the Middle Ages, beautifully hand-decorated to highlight the book's contents [3], up to visual symbols for computer-mediated communication of 21st century, such named Emoji [4], human beings established techniques to express knowledge and meaning through different visual representations.

Visual representations are a consistent component of human beings' evolution. They are a form of concept designs using combinations of colors, patterns, and geometrical figures [5].

Over the centuries, several forms of graphical visualization including drawings, photographs, diagrams, charts, tables, and many other creative forms [6], have been developed by artists, creative illustrators, scientists, designers, technologists, inventors through a mix of skills, technology, and available tools.

As history displays, graphical visualization has been used since early ages as a mean to transfer knowledge between human beings across generations. Recently, visual representations became recognized by various scientists as tools to ease cognitive comprehension in various scientific fields [7, 8].

According to Gilbert et al., graphical visualization is of great importance particularly in three aspects of the learning of science [9]. First, learning specific models that are currently used by a community of scientists such as the double-helix model of DNA, the P-N junction model of a transistor, or the visualization of a gene in biology. Second, learning to develop new qualitative models that explain an unexplored phenomenon by following a sequence of learning: to use an established model; to revise an established mod-el; to reconstruct an established model; to construct a model de novo. Third, learning to develop new quantitative models to make a comprehensive representation available. Therefore, visual representations are greatly important in science communication and education.

Given the incremental advance in the development of scientific research, which led to high growth rates in scientific publications [10], the importance of visualizing concepts, research outcomes, and results became more understood and appreciated among the scientific community. As a matter of fact, in the last decade, several editors encouraged the scientific community to create compelling, visually attractive, and clear graphical representations [11, 12, 13, 14].

Notwithstanding the general relevance of visual representations, there is not extensive recognition of the value of graphical representation to improve communication of scientific results and the level of awareness of visual design principles among non-design-centered communities.

The necessity to meticulously communicate complex information across several scientific fields, at a fast pace, is carrying the need to improve cross-disciplinary communicative graphical visualization strategies. Considering these aspects, this explorative study investigates the perception of STEM researchers, with initial limited or very low awareness about visual infographic design, and the value of visual representations as tools to support the communication of technical and scientific knowledge among academics and a wider non-design-centered community.

A user experience design researcher worked for 10 months with several researchers across the Aeronautics and Astronautics Department at Stanford University to infuse a "designer mindset" and to create visual representations and graphical visualizations to help improve the communication of research findings.

To measure the user perception and the value of visual representations in daily working practice, an interpretive approach was used, and a pilot study was conducted with a group of voluntary participants recruited among research members from the same department.

### 2. State of the art: Visual representations as a tool to improve communication

Historians assigned an important role in visual representation, particularly in early modern science [15]. Some of the most remarkable examples in this field are the extraordinary visual representation skills of Leonardo Da Vinci, an Italian polymath of the Re-naissance. The visual analysis of natural processes and technological developments created by the artist and inventor was transferred on paper through compelling graphical representations. These illustrations facilitated the communication of his thoughts and complex messages to his clients [15].

More specifically, Koyré pointed out that apart from the excellent artistic quality and detailed precision of Leonardo's anatomical drawings, the artist's major aim was to truly discover the inner structure and the mechanisms of the human body to make his knowledge available to a wider audience [16]. Leonardo believed that painting and visual representations, are not only superior to the other arts but can also be trusted to represent objects and concepts and understand them better than using only verbal descriptions [16].

A further example from the Renaissance comes from the extraordinary work of Galileo Galilei, an Italian astronomer, physicist, and engineer from the 17th century. He created a series of drawings illustrating the Earth's moon to demonstrate the power of his new invention — the telescope — to his patron, Cosimo de' Medici [17].

Visual representations have been widely used to facilitate the comprehension of com-plex information, such as the mechanics of a human body or a complex engineering system. Therefore, it is relevant to note that the combination of science, art, and visual de-sign makes progress and technical innovations more accessible and understandable to a wider audience.

In recent history, a variety of graphical languages and supporting tools were born. One of the most significant languages, that is still largely used for graphical representations, is the Vienna Method of Pictorial Statistics developed by Otto Neurath. Together with Marie Reidemeister and Gerd Arntz, Neurath created a new visual language to ex-plain complex systems by using pictorial representations. In 1934 it became known as the ISOTYPE - the International System of Typographic Picture Education [18].

By looking at the literature, it is clear that a common strategy accompanying Leonardo's, Galilei's and Neurath's work was to improve the communication of findings, inventions, or simply information in a more understandable manner for a wider audience. The need of sharing knowledge and disseminate the advances of scientific research is a crucial aspect that helps to improve and develop further research among the scientific community.

The massive quantity of information and correlations between several fields of scientific knowledge [19], as well as the need from readers to gain more information in less time as possible, are key aspects to consider nowadays.

It appears that in recent years, more and more researchers perceive the need to create compelling visual representations that convey complex arrays of data in colorful displays for their publications [20]. Scientists also rely on diagrams, figures, graphs, and abstracted visual representations when problem-solving to assimilate information, perceive trends, and conceptualize spatial relationships that relate their message to peers and colleagues across disciplines [21].

Abstraction is an important technique used in the creation and interpretation of scientific illustrations [22] and together with visual design principles [23, 24], they constitute some of the auxiliary tools that could help researchers to improve the communication of research findings. To understand and map the effectiveness of these tools as an optimal solution to support scientific research communication an exploratory study was carried on.

### 3. Research Methodology

This exploratory pilot study is grounded in the interpretive approach to user research, where feedback was elicited and analyzed to gain empathetic understanding and map the perception of researchers on visual design as an instrument to improve communication of research.

The research process and method were defined according to the literature review findings and after several explorative conversations to understand human actions on the relevance of the topic.
A user experience design researcher worked for 10 months with several academic staff members across the Aeronautics and Astronautics Department at Stanford University, collaborating to design visual representations and data visualizations as well as to infuse a "designer mindset" aiming to improve communication of scientific results of the research group.
To measure the impact of visual representations in daily working practice, a pilot study was dispensed to a group of voluntary participants recruited among faculty, post-doctoral, Ph.D. researchers, and graduate students from the Aeronautics and Astronautics Department at Stanford University.
The study, released in English language, proposed a list of closed and open-ended questions aiming to target four specific goals: a) to investigate the pre-collaboration level of awareness on visual design, b) to understand the impact of visual design in communicating research work, c) to evaluate the learning experience that participants developed across the collaboration, d) to understand the lesson learned on effective, clear visual design communication. Best practices in creating the study were followed [25], in particular avoiding ambiguous or double-barreled questions and questions containing double negatives.

### 3.1 Participant's recruitment

Participants recruitment was managed through an internal mailing system and an instant messaging platform normally used for communications within the group affiliates. A relatively small sample size of fourteen subjects (n=14) who were relevant to the problem under consideration was considered. Notwithstanding the limited number of participants, this exploratory study could give an initial contribution to establish the presence of one or multiple factors in the use of visual design principles for graphical representations in scientific research. Participants who were involved in experimenting with new visual illustration strategies and experienced benefits of communicating research outcomes through persuasive visual illustrations were considered.
Before a full release of the study to participants, a study verification was performed with a target group of researchers. This process allowed to identify (a) whether respondents clearly understood the questions and (b) whether questions had the same meaning for all respondents.
The study was conducted following ethical manners, under the best research practice of confidentiality and informed consent. The respondents' right to confidentiality as stated in the study description, together with a full description of the procedure, as well as the opportunity to drop out of the study at any time. The study had no time limit and participants were recommended to answer all questions. At the end of the study, participants didn't receive any compensation, and the average time for completion was six minutes.

### 3.2 Data collection

The pilot study was created using an open-source platform. It consisted of a brief introduction with aim of the study, followed by a set of demographic information questions, a set of twelve closed-ended questions, and one last open-ended question. A bipolar scaling method measuring either positive or negative response to a statement was used to score responses.
To facilitate the choice of answers a five-point Likert Scale with symmetry of categories from a midpoint with distinctly defined linguistic qualifiers was used to guarantee that attributes were observed. Certain questions were provided with a yes and no answer or multiple selections between items.
The study was further divided into five different sections.
Section (1) focused on collecting participants' demographic data (age, gender, ethnicity).

Section (2) asked questions regarding (a) familiarity with a visual design before the start of the collaboration, (b) the importance of visual design to improve research communication, (c) the potential of visual design to allow a wider community to understand complex information, and (d) how valuable was the visual design contribution to their research. Section (2) made use of ordinal data with variables ordered in categories.

Section (3) revealed a comparison between two different examples of an original design and redesigned infographic, and a series of questions aimed to capture participants' perspectives about visually compelling, clear communication of the selected examples. These questions were important to explore if participants gained an understanding of the differences between two different representation styles of the same research item. Section (4) questioned participants about preferred features they were considering when thinking about compelling visual design illustrations (with a multiple-choice question, based on design principles) and what constitutes a cluster of essential visual design elements for communicative graphical illustrations. Section (5) focused on understanding how to facilitate the collaboration between designers and scientists, in terms of easing the development process of graphical representations, as well as the method to facilitate the collaboration.

## 4. Results

In total fourteen subjects (n=14), ages ranged between 22 to 34 years old (M= 26.5), with one identified as female and thirteen as male, participated in the study. Their ethnicity was various and despite the small sample size, it covers several geographical areas of the world (see Table 1). The sample size and composition are strongly formed by male subjects. The statistical significance could be considered as a potential limitation of the study; however, this is an explorative pilot study that opens up future research paths. This aspect also allows reflecting on the composition of individuals working in general in STEM disciplines. A descriptive statistics approach was used to summarize and organize the characteristics of the data set.

**Table 1.** Participants sample and demographic data.

| Variable | | n | Age (years) |
|---|---|---|---|
| Sex | All | 14 | M 26.5 |
| | Female | 1 | 27 |
| | Male | 13 | M 26.5 |
| | Prefer not to say | 0 | 0 |
| Ethnicity | Asian/Pacific Islander/Asian Indian | 4 | |
| | White and Asian | 1 | |
| | White | 7 | |
| | Middle eastern/North African | 2 | |

In section (2) frequency was collected and specifically focused on understanding familiarity with the visual design before the start of the collaboration. Five participants (n=5) were slightly familiar and an additional five (n=5) were somewhat familiar with visual design principles.
Only one subject was not at all familiar and one extremely familiar, whereas only two participants (n=2) were moderately familiar.
Regarding the importance of visual design to improve the communication of research work, half of the subjects (n=7) recognized that visual design is important to improve research communication and an additional six of them (n=6) agreed that is very important. Only one participant expressed neutral feedback.

The majority of participants (n=10) strongly agreed that visual design can allow a wider community across different disciplines to benefit from scientific results, and rather a small minority (n=3) agreed. Only one subject answered with a neutral response. Detailed results are listed in table 2 below.

**Table 2.** Section 2 results.

| Question | Results | | | | |
|---|---|---|---|---|---|
| 1-Before the start of the collaboration, how familiar were you with visual design principles? | Not at all familiar | Slightly familiar | Somewhat familiar | Moderately familiar | Extremely familiar |
| | n=1 | n=5 | n=5 | n=2 | n=1 |
| 2-Based on your experience collaborating with the designer, how important do you consider visual design to improve the communication of your work? | Not at all important | Low importance | Neutral | Important | Very important |
| | n=0 | n=0 | n=1 | n=7 | n=6 |
| 3-Do you think visual design allows a wider community across different disciplines to benefit from your scientific results? | Strongly disagree | Disagree | Neither agree nor disagree | Agree | Strongly agree |
| | n=0 | n=0 | n=1 | n=3 | n=10 |

Section (3) of the study focused on understanding the capacity of participants to evaluate which illustration was elaborated following graphic design principles accepted by the scientific community [23, 24]. Two sets of illustrations, reported in image 1 below, were used to analyze participants' views.

**Image 1.** Illustrations 1 and 2 were submitted to participants. The above images (Image 1a and Image 2a) represent the original graphic elaborated, and the lower images (Image 1b and Image 2b) represent the updated through graphic design principles.

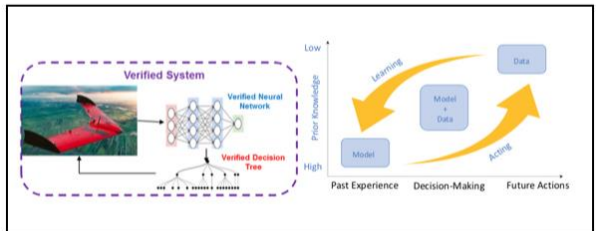
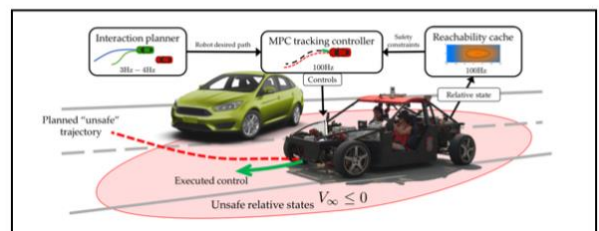
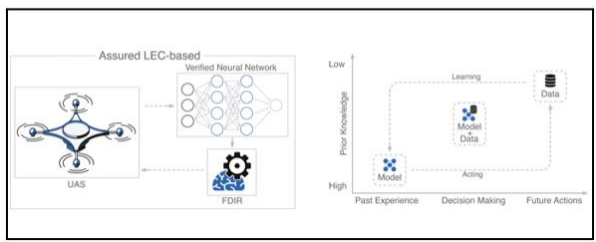
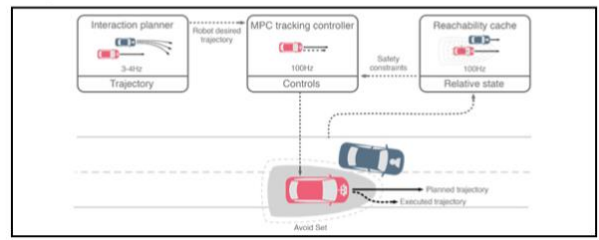

First, participants were asked to select which graphical representation offers a more visually compelling (e.g., widely understandable, catches the reader's attention) and clear (e.g., readable) communication. Regarding example 1, the majority of participants (n=10) chose the redesigned illustration (image 1b) and only a smaller number (n=4) selected the original illustration.

Considering example 2, almost all of the participants (n=13) chose the redesigned illustration (image 2b) and only one selected the original illustration.

Section (4) of the survey was built upon the examples previously shown and participants were queried to select from a list of eight visual design principles which were the four most important features they considered when expressed their previous preference on examples 1 and 2.

The eight visual design principles were namely: geometrical balance (equal distribution within the image), contrast (make elements stand out by emphasizing differences in size, color, direction), unity (similar elements appearing to belong together), dominance (have one element as the focal point and others being subordinate), visual minimalism (the tendency to simplify and organize complex visuals, by arranging the parts into an organized, minimal system), visual clarity (how a visual design can effectively prioritize and deliver information), color balance, and hierarchy between items.

The top four ranked were namely: visual minimalism (n=11), visual clarity (n=10), contrast and geometrical balance respectively with half of the preferences (n=7). Less participants selected in their top four design principles unity (n=6), hierarchy between items (n=6), color balance (n=5), and dominance (n=4).

The second question from section (4) focused on identifying what essential design elements people would consider when illustrating their research projects for well-organized communication. A list of eight elements (abundance of information, minimal information, mix-use of icons and text/numbers, use only text and numbers, geometric (uniform shapes), organic (non-uniform shapes), homogeneous visual language, non-homogeneous visual language) was proposed with a bipolar question format.

The table below emphasizes that most of the respondents have chosen to use minimal information (n=13), rather than the abundance of information, whereas fourteen participants were not favorable. The use of mixed icons, text, and numbers was selected by most of participants (n = 13), rather than only text and numbers, whilst fourteen participants were not favorable. In the case of geometric or organic shapes, the answers were less uniformed. Geometric shapes were the preferred choice for thirteen participants, whereas only five selected organic shapes.

A slightly similar result came out from the preference of homogeneous visual language, where thirteen respondents (n=13) agreed and eleven (n=11) disagreed on the use of non-homogeneous visual language.

**Table 3.** Section 4 results.

| Question | Yes | No |
|---|---|---|
| Abundance of information | n=0 | n=14 |
| Minimal information | n=13 | n=1 |
| Mix use of icons and text/numbers | n=13 | n=1 |
| Use only text and numbers | n=0 | n=14 |
| Geometric, uniform shapes, lines, colors | n=13 | n=1 |
| Organic, non-uniform shapes, lines, colors | n=5 | n=9 |
| Homogeneous visual language (similar icons, colors, font, size, geometry) | n=13 | n=1 |
| Non-Homogeneous visual language (different icons, colors, font, size, geometry) | n=3 | n=11 |

Section (5) of the survey mainly collected data on the collaboration between designer and scientists and how to maximize it in the future. In particular, subjects were asked to rank the three most desired options regarding the exchange of information between designer and scientist.

In a pre-pandemic time, the totality of participants expressed willingness to have face to face meetings (n=14). Almost all of them (n=13) also selected as second option a video call system, whereas half of them (n=7) found useful to share a folder for file exchange. Only five preferred an instant messaging system and two were favorable to an email exchange.

Notable of further consideration, these results would have been probably different if collected during or after the pandemic period. A further question explored how the best, more efficient communication of information between the scientist and the designer should happen. Three main options were available: one to one consultation (creation of an environment where the scientist explains its research to a non-expert (the design consultant) and several follow-ups with the scientist in a loop), group or team consultation (creation of an environment with the team of scientists explain their research to a non-expert (the design consultant) and several follow-ups from the team of scientists in a loop), multiple consultations (creation of an environment where a team leader scientist explains the research to a non-expert (the design consultant) and several follow-ups with members of the team in a loop).

Interestingly half of participants (n=7) preferred a one-to-one consultation as a mean to directly connect with the designer. Fewer subjects (n=5) opted for group or team consultation and only two were in favor of a multiple consultation approach.

## 5. Discussion

Research from Vogel et al. revealed that presentations using visual aids were found to be 43% more persuasive than unaided presentations [26].
In certain scientific domains, low awareness regarding the positive impact that visual design can give to improve research results communication still exists.
The early findings from this explorative study show that ahead of the collaboration between designers and scientists, a low or moderate tendency of familiarity with visual design happens between participants.
This statement was reinforced with a further question asked in section (3) of the study, that captured the capacity of participants to discern between an old-style graphic and a redesigned visual illustration, according to visual design principles. Furthermore, this section investigated which of the sample illustrations offered a more visually compelling and clear communication of the idea behind.
As the largest number of participants selected the redesigned version of the illustration both for example 1 (n=10) and example 2 (n=13), it is possible to identify that graphical representations created by following certain visual design principles [23, 24] offer a more visually compelling, clear communication.
These results were additionally verified by investigating which were the most important features that characterize a good visual design from the participants' perspective.
Interestingly, visual minimalism, which was repeatedly chosen by eleven participants (n=11), as well as visual clarity chosen by ten (n=10) were highly appreciated as understandable and perceivable design principles. Geometrical balance and contrast, unity, hierarchy between items, color balance, and dominance were instead selected less frequently.

This selection has the intention to highlight potentially less understanding about the application of those principles and therefore easiness to distinguish them in visual design. Follow up discussions with subjects posed the attention to several factors, such as the ability to recognize those principles, probably an easier task for a visual designer who is constantly working with such principles, or lack of awareness regarding the specific function or clarity about their application. The factors understood from this study underline that further investigation should be done.

Noteworthy results that confirm the tendency of participants to prefer certain visual design principles were further reported in section (4) of the study. What stood out was that following visual minimalism, all of respondents were against the use of excessive information and most of them were more likely to include a minimal amount of information in their visual illustrations.

In terms of visual clarity, the majority of participants preferred to use icons, with text and numbers and avoid the use of only text and numbers. These results confirm that visual minimalism and visual clarity are some of the fundamental aspects to take into account when creating visual designs.

More scattered results were reported in terms of geometric or organic shapes used for graphical illustrations. While a large preference was given for geometric illustrations, only a few participants would have chosen organic representations. This discrepancy in the result is probably given to the inner nature of the representation, personal taste, aesthetic, and context in which the representation will be published. Lastly, homogeneous visual language should be considered among different visual illustrations, in particular, the use of similar icons, colors, font, size, and geometry were recommended features.

To investigate the association between the importance of visual design to improve the communication of research work and the benefits that visual design can bring to an audience to better understand the research further analysis of questions number 2 and 3 (listed in table 2) was done.

Almost the totality of participants selected that is important (n=6) or very important (n=7) to consider visual design to improve the communication of the research work. Additionally, the majority (n=10) strongly agreed or agreed (n=3) that visual design can allow a wider community across different disciplines to benefit from the scientific achievements. The association of these two results pose some interesting remark on the importance of visual design and illustrations to improve communication and spread scientific results to a wider community across different disciplines.

An additional aspect to consider is the fluency in graphical visualization as a highly desirable skill to support a better understanding of concepts in scientific disciplines, in particular in the learning of science.

Final remarks focused particularly on considering how to best communicate between a design professional and a potential user which doesn't necessarily have a proper knowledge or awareness of visual design.

It is important to note that the study was developed and run just at the start of the pandemic in the USA (end of March 2020), so the answers were probably influenced by still a normal behavior at work, where face to face meetings and work from office was a daily regular practice.

Early findings show that all of the participants were keen on carrying on face-to-face meetings. That was a sign that the visual contact, the ability to sketch on a whiteboard, to reiterate thoughts and update the structure of the draft illustration in real-time, in-person was greatly appreciated.

The second most ranked answer emphasized the willingness to have online meetings through video-call systems. Despite a strong preference for face-to-face meetings, participants found extremely useful to operate the collaboration through online meeting systems, already in a pre-pandemic time.

A further aspect that could overcome the necessity of one-to-one in-person meetings, is the exchange of files, sketches, and material that is produced between designer and scientist. A shared folder for file exchange was backed as a reliable solution for half of the participants, whereas others preferred instant messaging systems or email exchange.

This partially answers on how the collaboration should be developed in a pandemic and eventually a post-pandemic era. We urge to say that repeating the study at a one-year distance, after the newly established working practice, would be helpful to understand if face-to-face meetings are essential. Furthermore, we believe that supplementary investigation should be done to more specifically comprehend the dynamics between designer and scientist, how ideas are developed during the discussion phase, and what the in-person meeting could convey in terms of creativity improvements [27].

This pilot study was of great importance to first provide an initial finding on the relevance that visual design has in strengthening the communication not only in business marketing or news information but heavily also in scientific fields such as STEM.

Second to understand how illustrations, following visual design principles, have the potential to allow a wider community, across different disciplines to better understand complex technical concepts. The image below (image 2) summarizes some of the major outcomes of this research.

**Image 2.** Infographic summary of main research findings.

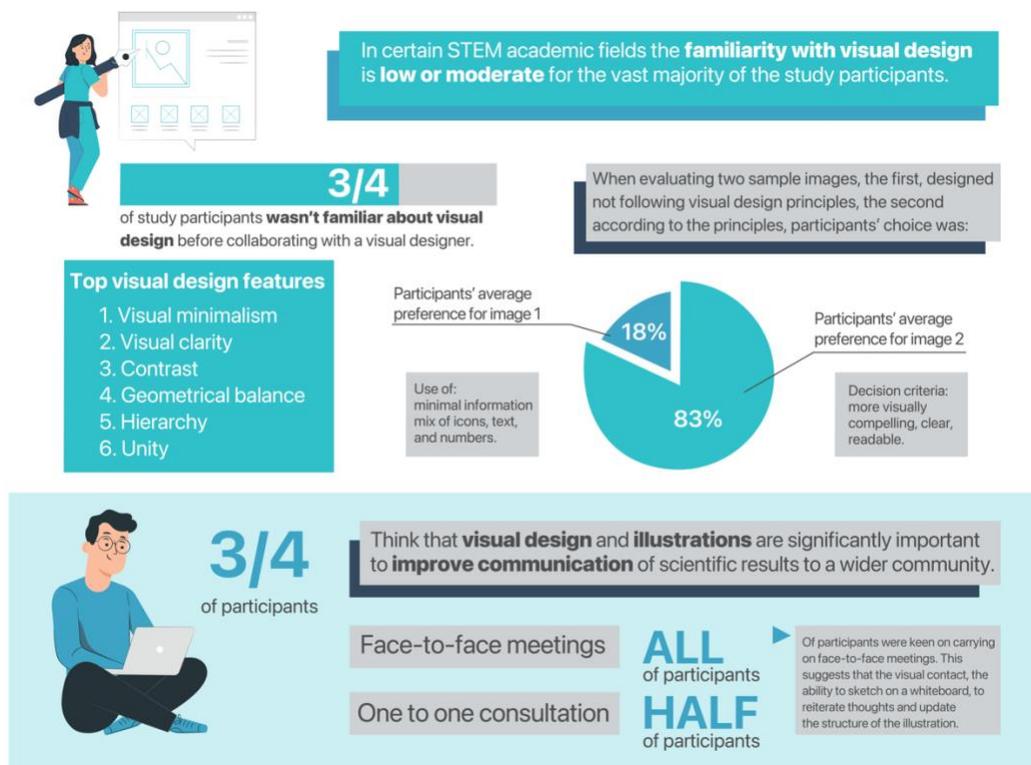

A further aspect that was noted in the study is the lack of awareness that STEM scientists, in particular, have about visual design and its principles. The collaboration between scientists and designers initially started as an aid to gain a better communication strategy of the research through visual representation, became later a training experience that allowed the research group to acquire new knowledge on visual design principles.

This finding, which wasn't originally planned as a major goal of the collaboration, poses the attention on the need of further expanding the education of non-visual designers about the potential and benefit that visual design could offer. The growing need to efficiently communicate complex information across several scientific fields is carrying the need to improve a cross-disciplinary visual design knowledge that will positively impact both the communicator and the receiver of the information. A comprehensive study organized with a larger sample size across several STEM disciplines, to further improve and consolidate the current research finding is needed. Additionally, it is suggested to target supplementary aspects such as the learning curve and experience of the non-design specialists, as well as discovering insights to develop further dissemination strategies to testify the importance of good design in visual communication.

## 6. Conclusion

Communication as well as receiving the information, are almost as important as the process of knowledge creation.
With this pilot study initial research was performed on visual design as a tool to improve the communication of research work. Visual design has the potential to provide an added value to communicate scientific results to a wider community across different disciplines.
Early evidence shows that visual representations positively support scientists to share research findings in a more compelling, visually clear, and impactful manner. Researchers, who included professionally executed visual illustrations in their publications, reports, and grant proposals, perceived their works as more easily readable by a wider community, with an increased impact as of its visual clarity and more likely to reach a wider audience across different disciplines.
Additional research shows that visual design has the power to interprets science and translate it into a concrete artifact [28]. This initial study confirms that improving the readability of information has the potential to foster the process of democratization of knowledge across disciplines, people, and cultures and allow a wider audience to benefit from the latest, cutting-edge scientific results.
As suggested by Rolandi et al. [24], a further aspect to consider for the future is to provide essential visual design foundations to improve the creation of visual illustrations for scientific research. With this research, we sought to contribute to strengthening the value that visual design can bring to scientific research and support early findings by Rolandi et al.
Fluency in graphical visualization can be considered a highly desirable skill to support a better communication of concepts in scientific disciplines, in particular in the learning of science. With relatively minimal improvements, visually compelling illustrations can considerably offer an enriched experience both to the reader, who clearly understands the concept and to the communicator, who can more easily transfer the knowledge to its audience.

## Acknowledgements

The collaboration and data collection were made possible thanks to the Department of Aeronautics and Astronautics and the Autonomous Systems Laboratory at Stanford University, USA. The data analysis and article development were made possible thanks to the funding from the European Union's Horizon 2020 research and innovation pro-gramme under the Marie Skłodowska-Curie grant agreement N° 846284, and the valuable advice from the members of the Engineering Design Centre at University of Cambridge, United Kingdom.